\documentclass[journal]{IEEEtran}
\usepackage{amsmath}
\usepackage{cite}
\usepackage{amssymb}
\usepackage{amsfonts}
\usepackage{algorithm}
\usepackage{algpseudocode}
\usepackage{dsfont}
\usepackage{graphicx}
\usepackage{epsfig}
\usepackage{subfigure}
\usepackage{psfrag}
\usepackage{xcolor}
\usepackage{url}
\usepackage[colorlinks,linkcolor=black,urlcolor=black,anchorcolor=black,citecolor=black,hyperfootnotes=true]{hyperref}
\usepackage{tikz}
\usepackage{stfloats}

\usepackage{enumitem}
\usepackage{setspace}

\newcommand{\mv}[1]{\mbox{\boldmath{$ #1 $}}}

\title{Covariance-Based Joint Device Activity and Delay Detection in Asynchronous mMTC}
\author{Zhaorui Wang, Ya-Feng Liu, and Liang Liu
\thanks{The work of Y.-F. Liu was supported in part by the National Natural Science Foundation of China (NSFC) under Grant 12022116 and Grant 12021001. The work of L. Liu was supported by the Research Grants Council, Hong Kong, China, under Grant 25215020. \emph{(Corresponding author: Ya-Feng Liu.)}}
\thanks{Z. Wang is with the Department of Information Engineering, The Chinese University of Hong Kong, Hong Kong (e-mail: zrwang2009@gmail.com).}
\thanks{Y.-F. Liu is with the State Key Laboratory of Scientific and Engineering Computing, Institute of Computational Mathematics and Scientific/Engineering Computing, Academy of Mathematics and Systems Science, Chinese Academy of Sciences, Beijing 100190, China (e-mail: yafliu@lsec.cc.ac.cn).}
\thanks{L. Liu is with the Department of Electronic and Information Engineering, The Hong Kong Polytechnic University, Hong Kong (e-mail: liang-eie.liu@polyu.edu.hk).}}

\begin{document}
\maketitle \thispagestyle{empty}

\begin{abstract}
In this letter, we study the joint device activity and delay detection problem in asynchronous massive machine-type communications (mMTC), where all active devices asynchronously transmit their preassigned preamble sequences to the base station (BS) for device identification and delay detection. We first formulate this joint detection problem as a maximum likelihood estimation problem, which depends on the received signal only through its sample covariance, and then propose efficient coordinate descent type of algorithms to solve the formulated problem. Our proposed covariance-based approach is sharply different from the existing compressed sensing (CS) approach for the same problem. Numerical results show that our proposed covariance-based approach significantly outperforms the CS approach in terms of the detection performance since our proposed approach can make better use of the BS antennas  than the CS approach.
\end{abstract}


\section{Introduction}\label{sec:Introduction}
This letter studies massive machine-type communications (mMTC) which provides efficient random access communications for a large number of devices, out of which only a small number of them are active\cite{bockelmann2016massive,chen2020massive}. To reduce  communication latency, a grant-free random access scheme was proposed in \cite{liu2018sparse,senel2018grant}, where the active devices can directly transmit the data signals after the preamble signals without permissions from the base station (BS). Note that in practice, since the IoT devices are equipped with low-cost local oscillators, they are not perfectly synchronized. This letter studies the problem of joint device activity and delay detection  in the asynchronous mMTC system. 

Most of the existing works on massive random access are based on the perfect synchronization assumption, see \cite{senel2018grant,liu2018massive,chen2018sparse,jiang2018joint,ke2020compressive,ding2019sparsity,sun2019exploiting,mei2021compressive,haghighatshoar2018improved,chen2019covariance,Chenphase,wang2021efficient,ganesan2020algorithm,chen2021sparse,accwang} and the references therein. There are generally two lines of researches for massive random access. One line of research is to  apply the compressed sensing (CS) approach to solve the joint activity detection and channel estimation problem by taking advantage of the sporadical nature of the device activity pattern \cite{senel2018grant,liu2018massive,chen2018sparse,jiang2018joint,ke2020compressive,ding2019sparsity,sun2019exploiting,mei2021compressive}. This line of approaches generally suffer from poor channel estimation performance (especially in massive MIMO) as shown in \cite{liu2018massive, liu2018massive2} since the channels are estimated through nonorthogonal preamble sequences. The second line of research is to apply the covariance-based approach\cite{haghighatshoar2018improved,chen2019covariance,Chenphase,wang2021efficient,ganesan2020algorithm,chen2021sparse,accwang} to solve the problem.
Specifically, if only device activity is detected, but channels are not estimated (e.g., a sensor to send a fire alarm whose activity itself is an alarming signal), the activity can be detected merely based on the covariance of the received preamble signals. The minimum preamble sequence length required in the covariance-based approach is much smaller than that of the CS approach for the same performance\cite{haghighatshoar2018improved,Chenphase}. However, in many applications, data needs to be transmitted after the preamble, and it is important to estimate the channels based on the preamble. In this case, \cite{kang2020minimum} proposed a new protocol. Specifically, at the first phase, the BS applies the covariance-based approach to detect device activity; at the second phase, the BS transmits a common feedback message to all the active devices to select orthogonal preambles; at the third phase, the BS estimates channels and detects data from the active devices based on the orthogonal preambles. In short, since the channels  are estimated through orthogonal preambles in the three-phase protocol\cite{kang2020minimum}, channel estimation performance should be much better than that  from the grant-free protocol\cite{liu2018sparse,senel2018grant} using the CS approach. 

All the above works are based on the perfect synchronization assumption. This is impractical since it is hard to perfectly synchronize all devices in mMTC in general. In this case, the unknown delays of the active devices should be estimated as well. However, the current detection framework based on the synchronous assumption is inapplicable since the sensing/signature matrix in all of these works is not consistent with the actual one with delays. Recently,  \cite{liu2021efficient} considered the more practical \emph{asynchronous} massive random access scheme, and applied the CS approach to solve the joint device activity, delay detection, and channel estimation problem. 

In this letter, we consider the asynchronous massive random access scheme, and follow the covariance-based three-phase communication protocol in \cite{kang2020minimum}.  We particularly focus on the joint activity and delay detection problem in the first phase of the three-phase protocol in \cite{kang2020minimum}, and propose to use the covariance-based approach to solve it. We emphasize that the delay detection is  necessary under the protocol proposed in \cite{kang2020minimum} since the knowledge of delay can be utilized in the rest of phases. In this case, a key challenge is how to efficiently estimate device activity and delay at the BS  based on the preassigned preambles. To solve the above challenge, we formulate the detection problem as a maximum likelihood estimation problem. Based on the sample covariance of the received signals, we first adopt the  coordinate descent (CD) algorithm\cite{haghighatshoar2018improved}, then propose a block coordinate descent (BCD) algorithm to solve the detection problem, respectively. Numerical results show that our proposed covariance-based approaches have significant performance improvement compared with the CS  approach in \cite{liu2021efficient}.  A key intuition is that, the covariance-based approach can make  better use of the BS antennas such that its solution is more accurate than that of the CS approach. It is worth mentioning that the techniques developed in this paper can be further extended to the cell-free massive connectivity\cite{ke2020massive} in which the devices are more difficult to synchronize than the single-cell setup studied in this paper.

\section{System Model and Problem Formulation}\label{sec:SYS}
\subsection{System Model}
This letter studies a narrowband uplink  system which consists of a BS equipped with $M$ antennas and $N$ single-antenna devices. Due to the sporadic traffic,  only $K\ll N$ devices are active at each coherence block. Let $\alpha_n$ be an activity indicator of device $n$, and 
\begin{equation}
\alpha_n=\left\{\begin{array} {ll} 1, ~ {\rm if ~ device} ~ n ~ {\rm is ~ active}, \\ 0, ~ {\rm otherwise}, \end{array} ~~n=1,\dots, N.\right. 
\end{equation}
The set containing the indices of  $K$ active devices is denoted as $\mathcal{K}=\{n: \alpha_n=1, \forall n\}$. We assume block-fading channels, i.e., the channels remain roughly constant within each coherence block, but may vary among different coherence blocks. Let $\sqrt{\beta_n}\mv{h}_n\in\mathbb{C}^{M\times 1}$ be the channel between device $n$ and the BS, where $\beta_n$ is the large-scale fading component, and $\mv{h}_n$ follows the independent and identically distributed (i.i.d.) Rayleigh fading model, i.e., $\mv{h}_n\sim\mathcal{CN}(0,\mv{I})$, $\forall n$. In addition, the devices are not perfectly synchronized since the devices are equipped with low-cost local oscillators. We assume that the transmitted packet of each active device $n\in \mathcal{K}$ is delayed by $\tau_n\in\{0,\dots,\tau_{{\rm max}}\}$ symbols, where $\tau_{{\rm max}}$ is the maximum delay for all the devices and is assumed to be known at the BS. The delay of each active device is unknown and needs to be estimated.

Each device is preassigned a unique preamble sequence $\mv{s}_n\in\mathbb{C}^{L\times 1}$, $n=1,\dots, N$, where $L<N$ is the preamble sequence length. Then, for active device $n\in\mathcal{K}$, it starts to transmit sequence $\mv{s}_n$ at the beginning of the $\tau_{n}+1$ time slot. Note that in this letter, one time slot equals one symbol duration. We use the two terms interchangeably. Given the delay $\tau_n$, define the effective preamble sequence of device $n$  as
\begin{align}
\bar{\mv{s}}_{n,\tau_n}=[\underbrace{0, \dots, 0}_{\tau_n}, \mv{s}_n, \underbrace{0, \dots, 0}_{\tau_{{\rm max}}-\tau_n}]^T, ~n=1,\dots,N.
\end{align}
In this case, the received signal $\mv{Y}\in\mathbb{C}^{(L+\tau_{{\rm max}})\times M}$ from time slot 1 to time slot $L+\tau_{{\rm max}}$ is expressed as
\begin{align}
\mv{Y}=\sum_{n=1}^{N}\alpha_n\sqrt{\beta_n}\bar{\mv{s}}_{n,\tau_n}\mv{h}^T_n+\mv{Z},\label{eq:rec}
\end{align}
where $\mv{Z}=[\mv{z}_1,\dots,\mv{z}_{M}]$
is the additive white Gaussian noise with $\mv{z}_m\sim \mathcal{CN}(0,\sigma^2\mv{I})$, $\forall m$. To express the received signal at the BS in a more compact form, denote 
\begin{equation}
\theta_{n,\tau}=\left\{\begin{array} {ll} 1, ~ {\rm if} ~ \alpha_n=1 ~ {\rm and}~\tau=\tau_n, \\ 0, ~ {\rm otherwise}, \end{array} ~\forall n, \tau.\label{eq:ind2}\right.
\end{equation}
In other words, $\theta_{n,\tau}=1$ only if device $n$ is active, and its delay is of  $\tau_n$ symbol durations. In addition, denote
$\mv{S}=\left[\bar{\mv{S}}_1,\bar{\mv{S}}_2, \dots, \bar{\mv{S}}_N\right]$, where $\bar{\mv{S}}_n=\left[\bar{\mv{s}}_{n,0},\bar{\mv{s}}_{n,1},\dots, \bar{\mv{s}}_{n,\tau_{{\rm max}}}\right]$; $\mv{\gamma}={\rm diag}\left(\mv{\gamma}_1,\mv{\gamma}_2,\dots,\mv{\gamma}_N\right)$, where $\mv{\gamma}_n={\rm diag}\left(\gamma_{n,0}, \dots, \gamma_{n,\tau_{{\rm max}}}\right)$ and $\gamma_{n,\tau}=\beta_n\theta_{n,\tau}$; $\mv{H}=\left[\mv{H}_1,\mv{H}_2,\dots, \mv{H}_N\right]^T$, where $\mv{H}_n=[\underbrace{\mv{h}_n,\mv{h}_n,\dots,\mv{h}_n}_{\tau_{{\rm max}}+1}]$. In this case, $\mv{Y}$ in \eqref{eq:rec} is rewritten as
\begin{align}
\mv{Y}=\sum_{n=1}^{N}\sum_{\tau=0}^{\tau_{{\rm max}}}\theta_{n,\tau}\sqrt{\beta_n}\bar{\mv{s}}_{n,\tau}\mv{h}^T_n+\mv{Z}=\mv{S}\mv{\gamma}^{\frac{1}{2}}\mv{H}+\mv{Z}. \label{eq:rec2}
\end{align}
From the above received signal expression in \eqref{eq:rec2}, we model the joint activity and delay detection problem as an effective sequence detection problem. Specifically, we detect the effective sequence through the detection of $\mv{\gamma}$, which contains the device activity and  delay information in its diagonal elements $\theta_{n,\tau}$'s. For example, a non-zero diagonal element $\gamma_{3,1}$ in $\mv{\gamma}$ indicates that the effective sequence $\bar{\mv{s}}_{3,1}\in\mv{S}$ is detected, which means device $3$ is active, and its delay is of one symbol duration. In this case,  to estimate the device activity and delay information, we can estimate  $\mv{\gamma}$ in \eqref{eq:rec2} directly, instead of estimating $\alpha_n$'s and $\tau_n$'s separately. 

\subsection{Problem Formulation}
In the following, we show the approach to estimate $\mv{\gamma}$ based on the received signal $\mv{Y}$ and the effective sequence $\mv{S}$ in \eqref{eq:rec2}. Once $\mv{\gamma}$ is estimated, we can determine the active devices and their corresponding delays based on the diagonal values in the estimated $\mv{\gamma}$. We formulate the estimation of $\mv{\gamma}$ as a maximum likelihood estimation problem. Specifically, we first derive the likelihood of $\mv{Y}$ given $\mv{\gamma}$ and $\mv{S}$  as
\begin{align}
p(\mv{Y}|\mv{\gamma},\mv{S})&=\prod_{m=1}^{M}p(\mv{y}_m|\mv{\gamma},\mv{S})\nonumber\\&\overset{(a)}{=}\frac{1}{|\mv{\Sigma}\pi|^M}\prod_{m=1}^{M}\exp{(-\mv{y}_m^H\mv{\Sigma}^{-1}\mv{y}_m)}\nonumber\\
&\overset{(b)}{=}\frac{1}{|\mv{\Sigma}\pi|^M}\exp{\left(-{\rm Tr}\left(\mv{\Sigma}^{-1}\mv{Y}\mv{Y}^H\right)\right)},\label{eq:li}
\end{align}
where $\mv{y}_m$ is the $m$-th column of $\mv{Y}$. In \eqref{eq:li}, (a) is because given $\mv{\gamma}$ and $\mv{S}$, $\mv{y}_m$'s are independent of each other, and (b) is because given $\mv{\gamma}$ and $\mv{S}$, $\mv{y}_m$ is Gaussian distributed with zero mean and covariance matrix
\begin{align}
\mv{\Sigma}=\mathbb{E}(\mv{y}_m\mv{y}_m^H)&=\mv{S}\mv{\gamma}^{\frac{1}{2}}{\rm diag}\left(\mv{E}_1,\dots,\mv{E}_N\right)\mv{\gamma}^{\frac{1}{2}}\mv{S}^H+\sigma^2\mv{I}\nonumber\\
&=\mv{S}\mv{\gamma}\mv{S}^H+\sigma^2\mv{I},\label{eq:sigma}
\end{align}
where $\mv{E}_n\in\mathbb{C}^{(\tau_{{\rm max}}+1)\times (\tau_{{\rm max}}+1)}$ is a all-one matrix. From \eqref{eq:sigma}, we know that the covariance matrix $\mv{\Sigma}$ does not depend on the index $m$.  Note that maximizing $p(\mv{Y}|\mv{\gamma},\mv{S})$ in \eqref{eq:li} is equivalent to minimizing $-\log{(p(\mv{Y}|\mv{\gamma},\mv{S}))}$. In this case, the optimization problem is formulated as
\begin{align}
&\underset{\mv{\gamma}}{\rm minimize} ~~~\log{|\mv{\Sigma}|}+{\rm Tr}\left(\mv{\Sigma}^{-1}\tilde{\mv{\Sigma}}\right)\label{eq:opt1}\\
&{\rm subject~to} ~~~\mv{\gamma}\ge 0,\label{eq:cons1_1}\\
& ~~~~~~~~~~~~~~~|\mv{\gamma}_n|_0\le 1,~ n=1,\dots, N,\label{eq:cons1_2}
\end{align}
where $\tilde{\mv{\Sigma}}=\frac{1}{M}\mv{Y}\mv{Y}^H$ is the sample covariance of the received signals;  $|\mv{\gamma}_n|_0$ denotes the number of nonzero elements in the vector $\mv{\gamma}_n$. The constraint in \eqref{eq:cons1_1} is due to the non-negative path loss coefficients in the diagonal elements of $\mv{\gamma}$, and the constraint in \eqref{eq:cons1_2} is because there is at most one possible delay for each device. In addition, the estimation problem depends on the received signals only through the sample covariance $\tilde{\mv{\Sigma}}$.  Note that, the sample covariance $\tilde{\mv{\Sigma}}$  approaches  the real covariance $\mv{\Sigma}$ as the increase of $M$, which in turn leads to a better solution. 

\begin{algorithm}
	\caption{CD Algorithm followed by Constraint Enforcement for Solving Problem \eqref{eq:opt1}}\label{alg:cd}
	\begin{algorithmic}[1] 
		\State Initialization: $\hat{\mv{\gamma}}=0$, $\hat{\mv{\Sigma}}=\sigma^2\mv{I}$, $f_0$, $\epsilon=1$, $i=0$.\label{line1}
		\While{$\epsilon>\delta$}
		\State $i\gets i+1$; 
		\For{$n=1,\dots, N$}	
		\For{$\tau=0,\dots, \tau_{{\rm max}}$}
		\State Compute $\eta$ according to \eqref{eq:alg1} on the next page; \label{line6}
		\State $\hat{\gamma}_{n,\tau}\gets\hat{\gamma}_{n,\tau}+\eta$; \label{line7}
		\State $\hat{\mv{\Sigma}}^{-1} \gets \hat{\mv{\Sigma}}^{-1}-\eta\frac{\hat{\mv{\Sigma}}^{-1}\mv{s}_{n,\tau}\mv{s}_{n,\tau}^H\hat{\mv{\Sigma}}^{-1}}{1+\eta\mv{s}_{n,\tau}^H\mv{\Sigma}^{-1}\mv{s}_{n,\tau}}$; \label{line8}
		\EndFor
		\EndFor
		\State Compute $f_{i}$;
		\State $\epsilon=f_{i-1}-f_i$;
		\EndWhile \Comment{\emph{End of the CD algorithm}.}
		\State Enforce constraint \eqref{eq:cons1_2} on $\hat{\mv{\gamma}}$ according to \eqref{eq:enf};
		\State Output $\hat{\mv{\gamma}}$. \label{line14}
	\end{algorithmic}
\end{algorithm}

\section{Proposed Algorithms for Joint device Activity and Delay Detection}\label{sec:method}
In this section, we propose two algorithms for solving the optimization problem in \eqref{eq:opt1}. Note that the optimization problem in \eqref{eq:opt1} is not a convex problem. We first adopt the standard CD algorithm\cite{haghighatshoar2018improved}, and then propose a BCD algorithm for solving the problem, respectively.

\subsection{CD Algorithm Followed by Constraint Enforcement}
The optimization problem \eqref{eq:opt1} can be efficiently solved by adopting the standard CD algorithm if we drop the constraint \eqref{eq:cons1_2} in the problem \eqref{eq:opt1}. In this subsection, we first apply the standard CD algorithm to get an estimate $\hat{\mv{\gamma}}$ of $\mv{\gamma}$. Then, we enforce the block-wise constraint in \eqref{eq:cons1_2} on the estimate $\hat{\mv{\gamma}}$ as follows:
\begin{align}
\hat{\gamma}_{n,\tau}=\left\{\begin{array} {ll} \!\!\!\!\hat{\gamma}_{n,\tau}, ~ {\rm if} ~\hat{\gamma}_{n,\tau}=\hat{\gamma}_{n}^{{\rm max}}, \\ 0, ~~~ {\rm otherwise}, \end{array}\!\forall n, \tau,\right. \label{eq:enf}
\end{align}
where $\hat{\gamma}_{n}^{{\rm max}}=\max_{\tau=0}^{\tau=\tau_{{\rm max}}}\{\hat{\gamma}_{n,\tau}\}$. 

The CD followed by constraint enforcement (CD-E) algorithm is summarized in Algorithm \ref{alg:cd},  where $i$ denotes the index of iteration, and $f_i$ denotes the objective  value of problem \eqref{eq:opt1} at the $i$-th iteration. In particular, $f_0$ denotes the objective  value when the initial estimation $\hat{\mv{\gamma}}$ is set to be zero; $\delta$ is a small number to control the convergence of the algorithm. After executing the CD-E algorithm, we get an estimate $\hat{\mv{\gamma}}$. Then, we need to do a thresholding procedure as follows:
\begin{align}
\hat{\gamma}_{n,\tau}=\left\{\begin{array} {ll} \!\!\!\!\hat{\gamma}_{n,\tau}, ~ {\rm if} ~  \hat{\gamma}_{n,\tau}\ge t_{{\rm CD}}, \\ 0, ~~~ {\rm otherwise}, \end{array}\!\forall n, \tau,\right. \label{eq:thd}
\end{align}
where $t_{{\rm CD}}$ is a threshold to determine the nonzero diagonal elements in $\hat{\mv{\gamma}}$ since the estimated $\hat{\mv{\gamma}}$ has been corrupted by noise. Last, the activity and delay indicator $\theta_{n,\tau}$ in \eqref{eq:ind2} is estimated as
\begin{equation}
\hat{\theta}_{n,\tau}=\left\{\begin{array} {ll} 1, ~ {\rm if} ~ \hat{\gamma}_{n,\tau}>0, \\ 0, ~ {\rm otherwise}, \end{array} ~~\forall n, \tau.\right. \label{eq:theta}
\end{equation}
The indices of nonzero $\hat{\theta}_{n,\tau}$ are the active devices and their corresponding delays.
 
\begin{figure*}[ht]
	\begin{equation}
	\eta=\max\left\{\frac{\mv{s}_{n,\tau}^H\hat{\mv{\Sigma}}^{-1}\tilde{\mv{\Sigma}}\hat{\mv{\Sigma}}^{-1}\mv{s}_{n,\tau}-\mv{s}_{n,\tau}^H\hat{\mv{\Sigma}}^{-1}\mv{s}_{n,\tau}}{\left(\mv{s}_{n,\tau}^H\hat{\mv{\Sigma}}^{-1}\mv{s}_{n,\tau}\right)^2}, -\hat{\gamma}_{n,\tau}\right\}\label{eq:alg1}
	\end{equation}
	\hrulefill
\end{figure*}

\subsection{Block Coordinate Descent Algorithm}
The CD-E algorithm presented in the last subsection applies the CD algorithm to optimize the problem \eqref{eq:opt1} by relaxing the constraint \eqref{eq:cons1_2}, and then enforces the obtained solution to satisfy the constraint in \eqref{eq:cons1_2}. In this subsection, we introduce a BCD algorithm which takes the constraint \eqref{eq:cons1_2} into account within its optimization process. Thus, the BCD algorithm  may lead to a better solution compared with the CD-E algorithm. 

For the BCD algorithm, at each time, we only optimize $\mv{\gamma}_{\bar{n}}$, while keeping $\mv{\gamma}_{n}$ for $n\ne \bar{n}$ unchanged. In this case, the optimization problem in \eqref{eq:opt1} reduces to the following problem: 
\begin{align}
&\underset{\mv{\gamma}_{\bar{n}}}{\rm minimize} ~~~\log{|\mv{\Sigma}|}+{\rm Tr}\left(\mv{\Sigma}^{-1}\tilde{\mv{\Sigma}}\right)\label{eq:opt2}\\
&{\rm subject~to} ~~~\mv{\gamma}_{\bar{n}}\ge 0,\\
& ~~~~~~~~~~~~~~~|\mv{\gamma}_{\bar{n}}|_0\le 1\label{eq:cons2}.
\end{align}
We propose to solve the problem \eqref{eq:opt2} by decomposing it into $\tau_{{\rm max}}+1$ subproblems. Specifically,  for the $\bar{\tau}$-th subproblem, we fix $\gamma_{\bar{n},\tau}=0$ for $\tau\ne\bar{\tau}$, and optimize $\gamma_{\bar{n},\bar{\tau}}$ by solving the following problem:
\begin{align}
&\underset{\gamma_{\bar{n},\bar{\tau}}}{\rm minimize} ~~~\log{|\mv{\Sigma}|}+{\rm Tr}\left(\mv{\Sigma}^{-1}\tilde{\mv{\Sigma}}\right)\label{eq:opt3}\\
&{\rm subject~to} ~~\gamma_{\bar{n},\bar{\tau}}\ge 0.
\end{align}	
The above problem \eqref{eq:opt3} is exactly a subproblem within the CD-E algorithm (i.e., Line \ref{line6} to Line \ref{line8} in Algorithm \ref{alg:cd}) and has a closed-form solution. For the $\bar{\tau}$-th subproblem, let $\hat{\gamma}_{\bar{n},\bar{\tau}}$ be the solution to the subproblem \eqref{eq:opt3}, $\mv{\gamma}_{\bar{n}}^{(\bar{\tau})}\in\mathbb{R}^{(\tau_{{\rm max}}+1)\times 1}$ be a all-zero vector except for the $\bar{\tau}$-th element being $\hat{\gamma}_{\bar{n},\bar{\tau}}$, and $f_{\bar{n},\bar{\tau}}$ be the objective value of problem \eqref{eq:opt3} at $\hat{\gamma}_{\bar{n},\bar{\tau}}$. After solving subproblem \eqref{eq:opt3} for $\bar{\tau}=0,\dots,\tau_{{\rm max}}$, the solution for problem \eqref{eq:opt2} is
\begin{align}
\mv{\gamma}_{\bar{n}}=\mv{\gamma}_{\bar{n}}^{(\hat{\tau})},~~ {\rm for}~~ \hat{\tau}=\underset{\tau}{\arg\min} ~ f_{\bar{n},\tau}\label{eq:alg2}.
\end{align}
The problem \eqref{eq:opt2} is solved from $\bar{n}=1$ to $\bar{n}=N$. The BCD algorithm is summarized in Algorithm \ref{alg:bcd}. After executing the BCD algorithm, we get an estimate $\hat{\mv{\gamma}}$. Then, we determine the nonzero diagonal elements in $\hat{\mv{\gamma}}$ according to \eqref{eq:thd} except that the threshold now is $t_{{\rm BCD}}$. Last, the activity and delay indicator $\theta_{n,\tau}$ in \eqref{eq:ind2} is similarly estimated as in \eqref{eq:theta}.

In Section \ref{sec:num}, we show that the BCD algorithm outperforms the CD-E algorithm. The possible reasons are as follows. First, BCD solves the original problem while CD-E solves the relaxation problem albeit with a constraint enforcement procedure. Second, the monotonicity of the objective values in  BCD is guaranteed. However, the constraint enforcement procedure in CD-E may destroy the monotonicity of the objective function, and possibly increases the objective value without any control.

\begin{algorithm}
	\caption{BCD Algorithm for Solving Problem \eqref{eq:opt1}}\label{alg:bcd}
	\begin{algorithmic}[1] 
		\State Initialization: $\hat{\mv{\gamma}}=0$, $f_0$, $\epsilon=1$, $i=0$.
		\While{$\epsilon>\delta$}
		\State $i\gets i+1$;
		\For{$\bar{n}=1,\dots, N$}	
		\For{$\bar{\tau}=0,\dots, \tau_{{\rm max}}$}
		\State Set $\hat{\gamma}_{\bar{n},\tau}=0$ for $\tau\ne\bar{\tau}$;
		\State Compute  $\eta$  according to \eqref{eq:alg1}; 
		\State $\hat{\gamma}_{\bar{n},\bar{\tau}}\gets\hat{\gamma}_{\bar{n},\bar{\tau}}+\eta$;
		\State Compute $f_{\bar{n}, \bar{\tau}}$;
		\EndFor
		\State Determine $\mv{\gamma}_{\bar{n}}$ according to \eqref{eq:alg2};
		\EndFor
		\State Compute $f_{i}$;
		\State $\epsilon=f_{i-1}-f_i$;
		\EndWhile
		\State Output $\hat{\mv{\gamma}}$.
	\end{algorithmic}
\end{algorithm}	
\vspace{-0.3cm}
\section{Numerical Results}\label{sec:num}
In this section, we present the detection performance of the proposed CD-E  and BCD algorithms in terms of missed detection probability (MDP) and false alarm probability (FAP). The MDP is the probability that an active device is detected to be inactive, or an active device is correctly detected but its delay is not correctly detected; the FAP is the probability that an inactive device is  detected to be active. The number of potential devices is $N=200$, the number of active devices is $K=90$, the preamble sequence length is $L=100$, and the maximum number of delays is $\tau_{{\rm max}}=4$ symbol durations. All preamble sequences follow the i.i.d. complex Gaussian distribution with zero mean and unit variance. The noise power spectrum density is -169dBm/Hz, the bandwidth is 10 MHz, the transmit power of each device is set to be 23dBm, and  the path loss is modeled as  $128.1+37.6\log_{10}(d)$, where $d$ is the distance between the BS and devices in km\cite{3GPP}. For ease of demonstration, we consider a worst case in which all devices are located in the cell edge such that the large-scale fading coefficients are the same for all devices. In the simulations, we set $d=1$ km.   Moreover,  we set $\delta=10^{-3}$, $t_{{\rm CD}}=0.1$, and $t_{{\rm BCD}}=0.12$. Note that  $t_{{\rm BCD}}$ is slightly larger than $t_{{\rm CD}}$ since we observed that the estimated $\hat{\mv{\gamma}}$ by the BCD algorithm is slightly larger than that by the CD-E algorithm. Last, we have two benchmarks to evaluate the performance of the proposed algorithms as follows:
\begin{itemize}
    \item \textbf{CS approach in \cite{liu2021efficient}}. Ref. \cite{liu2021efficient} also studied the joint device activity and delay detection problem. Different from us,  \cite{liu2021efficient} solved this problem by applying a CS approach.  
	\item \textbf{CD-E with $\tau_{{\rm max}}=0$ and $L=104$}. We consider a case in which the delay is zero, and the sequence length is $L=104$ which is the same as the effective sequence length of the case with $\tau_{{\rm max}}=4$ and $L=100$. In this case, without delay, the CD-E algorithm only needs to detect the device activity, and the CD-E algorithm does not need to execute the constraint enforcement procedure in \eqref{eq:enf}. Thus, CD-E with $\tau_{{\rm max}}=0$ and $L=104$ provides a performance upper bound of  CD-E with $\tau_{{\rm max}}=4$ and $L=100$. In addition, the BCD algorithm is the same as the CD-E algorithm when $\tau_{{\rm max}}=0$.
\end{itemize}

\begin{figure}[t]
	\centering
	\includegraphics[scale=0.45]{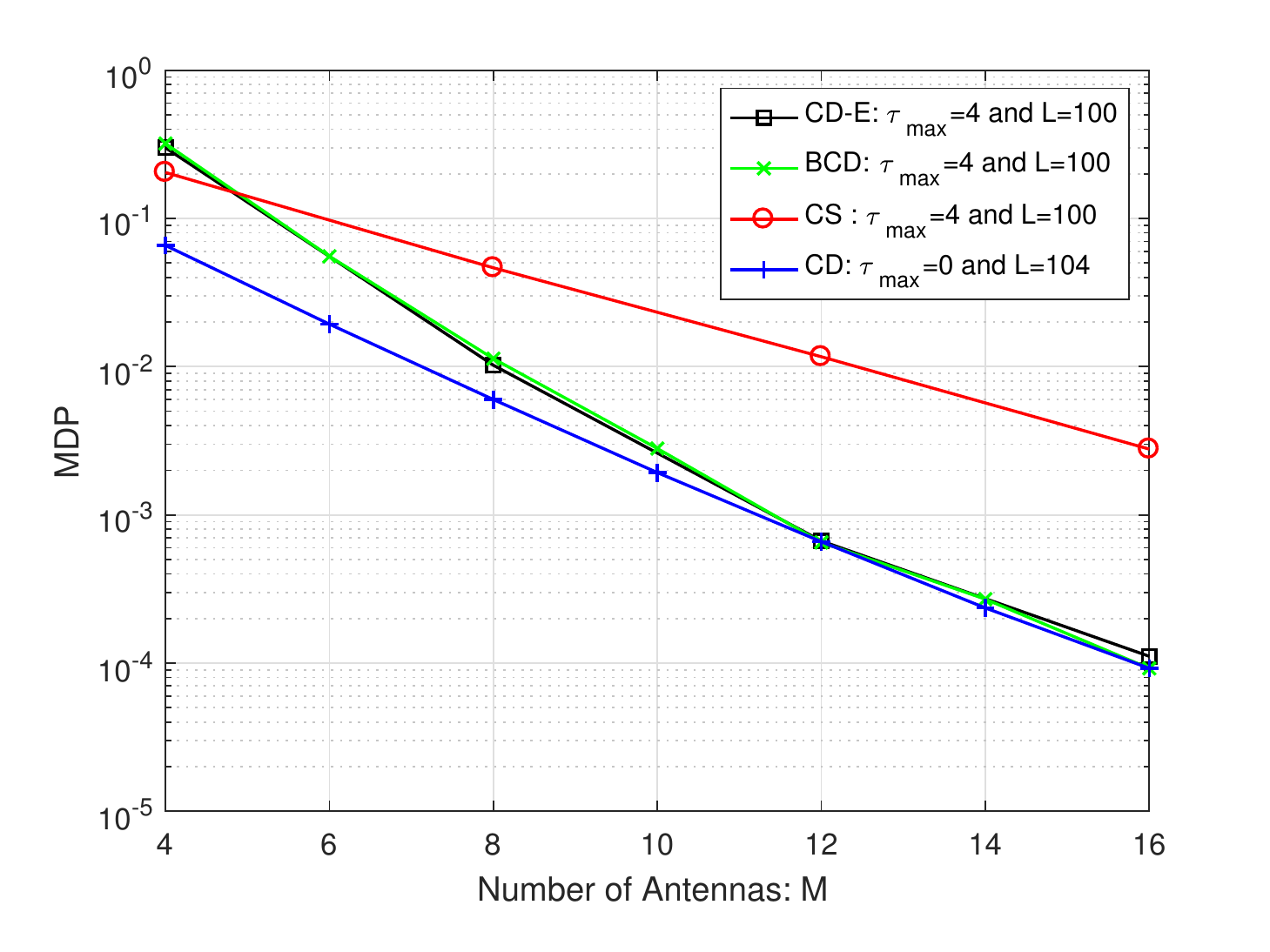}\\
	\caption{MDP comparison over different numbers of antennas.}\label{FigMDP}
\end{figure}
\begin{figure}[t]
	\centering
	\includegraphics[scale=0.45]{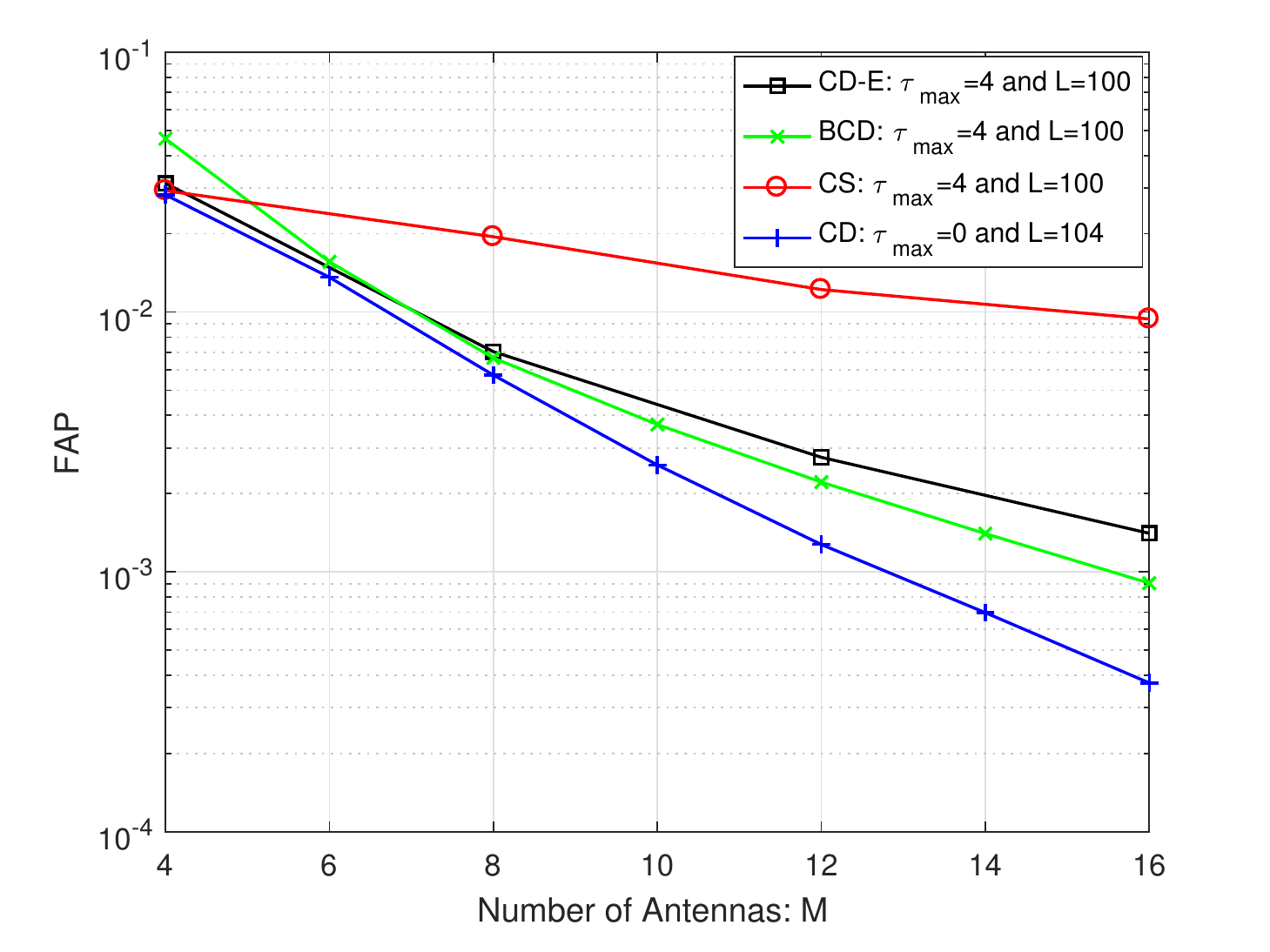}\\
	\caption{FAP comparison over different numbers of antennas.}\vspace{-0.5cm}\label{FigFAP}
\end{figure}

We compare their performance in terms of MDP and FAP over different numbers of BS antennas in Fig. \ref{FigMDP} and Fig. \ref{FigFAP}, respectively. First, when $\tau_{{\rm max}}=4$ and $L=100$,  CD-E and BCD  outperform CS significantly both in MDP and FAP. A key intuition is that, the covariance-based approach can make better use of the BS antennas such that the sample covariance of the received signals approaches the real covariance as the increase of the number of BS antennas,  which in turn leads to a better solution. Second, when $\tau_{{\rm max}}=4$ and $L=100$, BCD performs slightly better than CD-E. Specifically, from Fig. \ref{FigMDP}, they have the same MDP performance, while from Fig. \ref{FigFAP}, BCD performs better than CD-E in FAP when $M$ is larger than 8.  Third, in terms of MDP, ``CD-E/BCD: $\tau_{{\rm max}}=4$ and $L=100$'' perform roughly the same as ``CD-E: $\tau_{{\rm max}}=0$ and $L=104$'', which is the performance limit of ``CD-E/BCD: $\tau_{{\rm max}}=4$ and $L=100$''; in terms of FDP, the performance gap between ``CD-E/BCD: $\tau_{{\rm max}}=4$ and $L=100$'' and ``CD-E: $\tau_{{\rm max}}=0$ and $L=104$'' is  small. 

%\section{Conclusion}
%We studied asynchronous mMTC in which devices are sporadically active. Each active device transmits a preassigned sequence to the BS for device identification and delay detection. To solve this detection problem, we first modeled the joint device activity and delay detection problem as a sequence detection problem. Then we formulated the detection problem as a maximum likelihood estimation problem. Last, based on the covariance of the received signals, we proposed CD-E and BCD algorithms to solve the detection problem, respectively. Numerical results show that, our proposed covariance-based algorithms have significantly performance improvement compared with that of the compressed sensing approach since the covariance-based approach can make better use of the BS antennas to improve its detection accuracy.
\bibliographystyle{IEEEtran}
\bibliography{CIC}

\end{document}